\documentstyle[preprint,aps,epsfig]{revtex}

\draft
\begin{document}
\newcommand{\fiv}{\frac{1}{T_{1}}\phi^{m}_{T_{1}}(v)}
\newcommand{\Fiv}{\frac{1}{T_{2}}\phi^{m}_{T_{2}}(v)}
\newcommand{\fiV}{\frac{1}{T_{1}}\phi^{m}_{T_{1}}(V)}
\newcommand{\FiV}{\frac{1}{T_{2}}\phi^{m}_{T_{2}}(V)}
\newcommand{\ilV}{\int_{-\infty}^{V}dv}
\newcommand{\igV}{\int_{V}^{\infty}dv}

\title{From the adiabatic piston to macroscopic motion\\
induced by fluctuations}
\author{J.\ Piasecki}
\address{Institute of Theoretical Physics, University of Warsaw,
Ho\.{z}a 69, PL-00 681 Warsaw, Poland}
\author{Ch.\ Gruber}
\address{Institut de Physique Th\'{e}orique, Ecole Polytechnique
F\'ed\'erale de Lausanne, CH-1015 Lausanne, Switzerland}

\date{\today}

\maketitle
\begin{abstract}
\baselineskip 0.75 cm
The controversial problem of the evolution of an isolated system with an
internal adiabatic wall is investigated with the use of a simple
microscopic model and the Boltzmann equation. In the case of two infinite
volume
one-dimensional ideal fluids separated by a piston whose mass is equal to
the mass of the fluid particles we obtain a rigorous explicit stationary
non-equilibrium solution of the Boltzmann equation. It is shown that
at equal pressures on both sides of the piston, the temperature difference
induces a non-zero average velocity, oriented toward the region of higher
temperature. It thus turns out that despite the absence of macroscopic forces
the asymmetry of fluctuations results in a systematic macroscopic motion.
This remarkable effect is analogous to the dynamics of stochastic ratchets,
where fluctuations conspire with spatial  anisotropy to generate directed
motion. However, a different mechanism is involved here. The relevance of
the discovered motion to the adiabatic piston problem is discussed.
\end{abstract}


\section{Introduction}
The recent work of E.Lieb and J.Yngvason \cite{lieb98} on the foundations
of thermostatics has given a new interest to an old but still controversial
problem of thermodynamics. This is the so called "adiabatic piston
problem" described as follows \cite{callen63} (see also recent references
in \cite{gruber98}). The system is composed of a finite volume cylinder
containing two fluids separated by an adiabatic movable piston, as shown
in Fig.1. A brake maintains initially the piston at rest, and the two
fluids are in equilibrium with temperature, pressure and volume given by
$(T_{1}^{(0)}, p_{1}^{(0)}, V_{1}^{(0)}=A(L_{1}+X^{(0)}))$, and
$(T_{2}^{(0)}, p_{2}^{(0)}, V_{2}^{(0)}=A(L_{2}-X^{(0)}))$, respectively.
$A$ denotes here the cross section area of the cylinder, and $X^{(0)}$ the
initial position of the piston. At a certain time the brake is released,
and the problem is to find the final equilibrium state.

\begin{center}
\epsfig{file=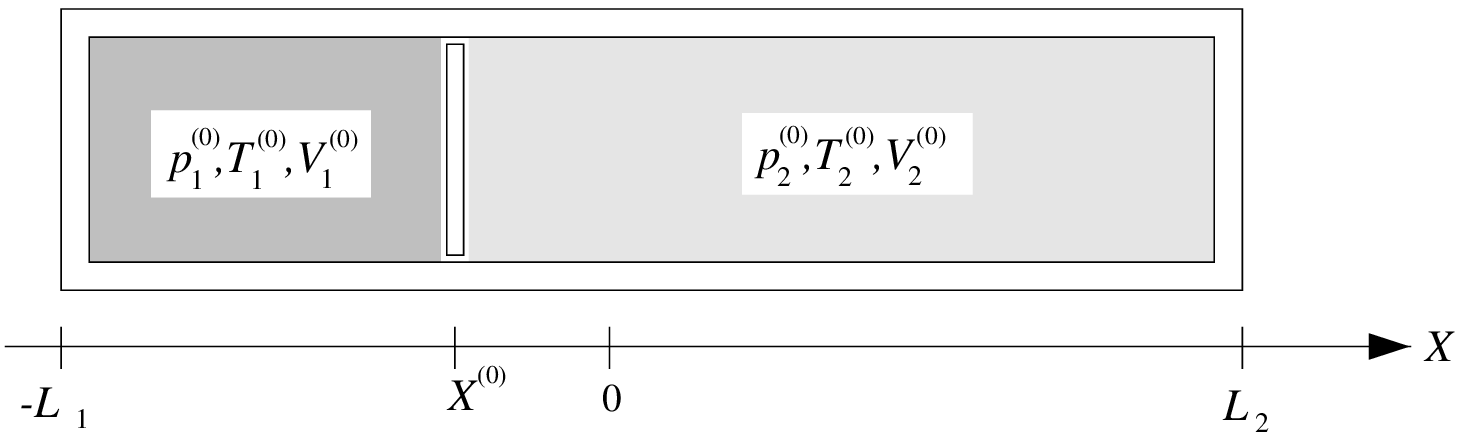} \\
Figure 1: Adiabatic piston problem
\end{center}

The laws of thermostatics give immediately the condition for the
mechanical equilibrium, namely that in the final equilibrium state the
pressures must be equal, i.e. $p_{1}^{f}=p_{2}^{f}$ (with $E_{1}^{f}
+ E_{2}^{f}=E_{1}^{0} + E_{2}^{0} $). However, they give no information
concerning the final temperatures, except for the condition that they
must be such that for both fluids the final entropy cannot be
smaller than its initial value, i.e. $S_{1}^{f} \geq S_{1}^{0}$, and
$S_{2}^{f} \geq S_{2}^{0}$.

On the other hand, from the laws of thermodynamics one can derive ordinary
differential equations for the time evolution of the system. Solving these
equations would then give the unique solution of the problem, and in
particular the final temperatures. This idea has been recently followed in
\cite{gruber98}, where a system of thermodynamic evolution equations has
been derived within a very simple model. One had to distinguish then between
the case where the two fluids have vanishing viscosities and the case where
at least one of the fluids has a non-zero viscosity. For non-viscous fluids the
thermodynamic equations predicts a periodic motion subject to the
energy conservation law, with constant subsystem entropies:
$S_{1}(t)=S_{1}^{(0)}$, $S_{2}(t)=S_{2}^{(0)}$.
When at least one of the fluids is viscous,
the  system will attain the final equilibrium state
with equal pressures $p_{1}=p_{2}$. However, to find the final entropies
(or temperatures) one has to solve the evolution
equations which so far could not be achieved explicitly. These same evolution
equations were also obtained in the framework of the kinetic theory of gases, assuming the fluctuations of the piston to be negligibly small and considering perfect gases, and several numerical solutions have been obtained \cite{crosignani96}.

But even having obtained the macroscopic description of the evolution of the
system the controversial question remains whether the stochastic frictionless
motion of the piston induced by collisions with the fluid particles could
eventually lead (on a much longer time scale) to a final state
where the temperatures of the fluids are equal, and whether such
a microscopic fluctuation mechanism would not violate the second law of
thermodynamics. Clearly, the analysis of such problem requires the
microscopic description of the dynamical evolution.

In this article we thus present a microscopic study of the motion of the
adiabatic piston in contact with two infinite volume reservoirs. Our
analysis, based on the Boltzmann equation, has been inspired by the above
fundamental adiabatic piston question.

In Section 2 we define the microscopic model and we discuss the applicability
of the Boltzmann equation to the description of the motion of the piston.
The rigorous determination of the stationary state in one dimension for the
case where the mass of the piston is equal to the fluid particle mass is
presented in Section 3. The derived solution at equal temperatures
reduces simply to the Maxwell equilibrium state. But when the temperatures are
different a non-equilibrium  stationary state appears characterized by
an average velocity of the piston oriented toward the higher temperature
region, although no macroscopic force is present (the pressures
on both sides of the piston are supposed to be equal). This situation
is thus precisely that of the ``molecular motor problem''. The rigorous
derivation (within Boltzmann's theory) of the fluctuation induced
macroscopic  motion is the main result of our work.
The relevance of this prediction to a much more realistic problem of a
massive piston is discussed in the closing Section 4. The case where the
mass of the piston is much larger than that of a fluid particle will be
discussed in the forthcoming paper \cite{gruber99}.

\section{Adiabatic piston separating two thermostats: the Boltzmann equation}

In order to simplify the analysis as much as possible we consider in the
following the one-dimensional system occupying the whole infinite line.
The two fluids are assumed to be perfect gases composed of identical point
particles of mass $m$. They are separated by a structureless particle of
mass $M$ which plays the role of the piston.
The perfectly elastic collisions between the
fluid particles result in instantaneous exchanges of velocities. So,
from the point of view of the dynamics of the piston the fluid particles can
be considered as non-interacting, penetrating each other in their unperturbed
free motion. The only perturbation of the fluid states is due to
collisions with the piston. The similar model has been already
investigated almost forty years ago by Lebowitz \cite{lebowitz59} and few
years later by Alkemade, van Kampen and Mac Donald \cite{vankampen}.

When a collision occurs, the pre-collisional velocities $V$ and $v$ of
the piston and the fluid particle, respectively, suffer a change according to
the elastic collision law
\begin{eqnarray}
V & \rightarrow & V' = V + \frac{2m}{M+m}(v-V) \label{1} \\
v & \rightarrow & v' = v - \frac{2M}{M+m}(v-V) \nonumber
\end{eqnarray}
Let us suppose that at the initial moment the fluids to the left and to the
right of the piston are at equilibrium with temperatures $T_{1}$ and
$T_{2}$, and number densities $n_{1}$ and $n_{2}$, respectively. Moreover,
the macroscopic mechanical equilibrium will be assumed: the pressures on
both sides of the piston are equal

\begin{equation}
p_{1}= n_{1}k_{B}T_{1} \; = \; p_{2}= n_{2}k_{B}T_{2} \label{2}
\end{equation}
($k_{B}$ is Boltzmann's constant).

When $M \gg m$, the probability that the piston interacts back with the fluid
particle it already collided with becomes negligible. In the absence of
recollisions the piston always "sees" the unperturbed Maxwell distributions
of the arriving fluid particles, and no correlations can occur between its
velocity and the precollisional velocity of the fluid particles. The
perturbations caused by collisions with the piston go away to infinity.
 So, in the large mass limit
$m/M\to 0$, we can expect the linear Boltzmann equation to give a correct
 description of the dynamical evolution of the piston. In order to write down
this equation we denote by $\phi^{m}_{T_{1}}$ and $\phi^{m}_{T_{2}}$
the Maxwell velocity distributions of the two surrounding fluids, where
\begin{equation}
\phi^{m}_{T} = \sqrt{\frac{m}{2\pi k_{B}T}} {\rm exp}\left( -\frac{mv^{2}}
{2k_{B}T} \right) \label{3}
\end{equation}
Let $F(X,V;t)$ denote the probability density for finding the piston at
point $X$ with velocity $V$ at time $t$. When $m/M \ll 1$, $F(X,V;t)$
satisfies the linear Boltzmann equation of the form
\begin{eqnarray}
{} & {} & \left( \frac{\partial}{\partial t}+ V\frac{\partial}{\partial X}
\right)F(X,V;t) =  \int dv |V-v|\times  \label{4} \\
{} & {} & [ \theta (V-v)n_{1}\phi^{m}_{T_{1}} \left( v-\frac{2M}{M+m}
(v-V)\right)F\left( X,V+\frac{2m}{M+m}(v-V);t \right) \nonumber \\
{} & {} & -\theta (v-V)n_{1}\phi^{m}_{T_{1}} (v) F(X,V;t) \nonumber \\
{} & {} & +\theta (v-V)n_{2}\phi^{m}_{T_{2}} \left(v-
\frac{2M}{M+m}(v-V)\right)F\left( X,V+\frac{2m}{M+m}(v-V);t \right)
\nonumber \\
{} & {} & -\theta (V-v)n_{2}\phi^{m}_{T_{2}} (v) F(X,V;t) ] \nonumber
\end{eqnarray}
where $\theta$ is the unit Heaviside step function.

Our aim is to determine the velocity distribution of the piston in the
asymptotic stationary state attained for $t\to \infty $. It turns out that
this problem can be given a rigorous answer in the special case of $M=m$.
But when the mass $M$ of the piston is equal to the fluid particle mass
$m$, the recollisions can take place, and the validity of the
Boltzmann equation becomes questionable. However, what we are really
interested in is the case of $M\gg m$. We thus expect that neglecting the
effect of recollisions in the equal mass case will lead to the evolution
qualitatively similar to that occuring for  $M\gg m$, simply because the
same class of dynamical events will be taken into account. This should
apply in particular to the nature of the stationary state.
It should be noticed here that even when compared with the exact results
the linear Boltzmann equation for $M=m$
represented a very reasonable approximation to the velocity correlation
function and to the Van Hove self-correlation function of the one-dimensional
hard rod fluid \cite{resibois78}.

Having adopted the above point of view we are going to study in the next
section the equation satisfied by the stationary probability density
$\Phi (V)$ for finding the piston with velocity $V$ in the equal pressure
case (see eq.(\ref{2})). Denoting by $F(X,V)$ the stationary solution of
equation (\ref{4}), we have
\begin{equation}
\Phi (V) = \int dX F(X,V)  \label{5}
\end{equation}

\section{Solving the Boltzmann equation: appearance of a molecular motor}

According to the linear Boltzmann equation (\ref{4}) the stationary velocity
distribution $\Phi (V)$ in the equal pressure case (\ref{2}) satisfies the
equation

\begin{equation}
\int dv |V-v| \left[ \theta (v-V)\fiv + \theta (V-v)\Fiv \right]\Phi (V)
\label{6}
\end{equation}
\[ = \int dv |V-v|\left[ \theta (V-v)\fiV + \theta (v-V)\FiV \right]\Phi (v)\]

In the following we shall assume $T_{2}>T_{1}$.

It is convenient to express eq.(\ref{6}) in terms of the following two
functions

\begin{eqnarray}
I(V) & = & \int_{V}^{\infty}dv (v-V)\fiv + \int_{-\infty}^{V}dv (V-v)\Fiv
\label{7} \\
G(V) & = & \int_{V}^{\infty}dv (v-V)\Phi (v) + \int_{-\infty}^{V}dv (V-v)
\Phi (v) \label{8}
\end{eqnarray}

Using the identities
\[ |V-v|\theta (V-v) = \frac{1}{2}[(V-v)+|V-v|] \]
\[ |V-v|\theta (v-V) = \frac{1}{2}[-(V-v)+|V-v|] \]
we rewrite eq.(\ref{6}) in the form
\begin{equation}
I(V)\Phi (V) - \frac{1}{2}[\fiV + \FiV ]G(V)=
\frac{1}{2} (V-<V>)[\fiV - \FiV ] \label{9}
\end{equation}
where $<V>$ denotes the average velocity of the piston
\begin{equation}
<V> = \int dV \Phi (V) V \label{10}
\end{equation}
We now notice that the first- and second-order derivatives of the functions
$I(V)$ and $G(V)$ are given by
\begin{eqnarray}
I'(V) & = & -\int_{V}^{\infty}dv\fiv + \int_{-\infty}^{V}dv\Fiv \label{11} \\
I''(V) & = & \fiv + \Fiv \label{12} \\
G'(V) & = & -\int_{V}^{\infty} dv \Phi (v) + \int_{-\infty}^{V} dv \Phi (v)
\label{13} \\
G''(V) & = & 2\Phi (V) \label{14}
\end{eqnarray}
Therefore (\ref{9}) takes the form of a simple differential equation
\begin{equation}
I(V)G''(V) - I''(V)G(V) = [ I(V)G'(V) - I'(V)G(V) ]' \label{15}
\end{equation}
\[  = (V-<V>)[\fiV - \FiV ] \]
Using the asymptotic formulae
\begin{equation}
I(V) \rightarrow \left\{
\begin{array}{ll}
\phantom{-}\frac{1}{T_{2}}V, & {\rm for\ } V\to +\infty\\
-\frac{1}{T_{1}}V, & {\rm for\ } V\to -\infty
\end{array}
\right. \label{16}
\end{equation}

\begin{equation}
G(V) \rightarrow \left\{
\begin{array}{ll}
\phantom{-}(V-<V>), & {\rm for\ } V\to +\infty\\
-(V-<V>), & {\rm for\ } V\to -\infty
\end{array}
\right. \label{17}
\end{equation}
we can readily evaluate the limit

\begin{equation}
\lim_{V\to \pm \infty} [ I(V)G'(V) - I'(V)G(V) ] = \left(\frac{1}{T_{1}}
+ \frac{1}{T_{2}}\right)<V> \label{18}
\end{equation}

Integrating equation (\ref{15}) with the boundary condition (\ref{18}) yields
the relation
\begin{eqnarray}
I(V)G'(V) - I'(V)G(V) & = & \frac{k_{B}}{m} [\phi^{m}_{T_{2}}(V) -
\phi^{m}_{T_{1}}(V)] \label{19} \\
{} & + & <V>[\int_{V}^{\infty} dv\fiv + \int_{-\infty}^{V} dv\Fiv] \nonumber
\end{eqnarray}

We thus arrived at a first order differential equation for the unknown
function $G$. The solution of eq.(\ref{19}) can be now constructed by standard
methods. Introducing the auxiliary function $H(V)$

\begin{equation}
G(V) = I(V) H(V), \label{20}
\end{equation}

and using the relations

\begin{eqnarray}
\lim_{V\to +\infty} H(V) & = & \lim_{V\to +\infty}\frac{G(V)}{I(V)} = T_{2}
\label{21} \\
\lim_{V\to -\infty} H(V) & = & \lim_{V\to -\infty}\frac{G(V)}{I(V)} = T_{1}
\nonumber
\end{eqnarray}

one readily determines the function $H$. It reads

\begin{eqnarray}
2H(V) & = & T_{1}+T_{2} + \int \frac{du}{I^{2}(u)}{\rm sgn}(V-u)\frac{k_{B}}
{m}[\phi_{T_{2}}^{m}(u)-\phi_{T_{1}}^{m}(u)] \label{22} \\
{} & + & \int \frac{du}{I^{2}(u)}{\rm sgn}(V-u)\left[ \int_{-\infty}^{u}dv\Fiv
 + \int_{u}^{\infty}dv\fiv \right]<V> \nonumber
\end{eqnarray}

Here ${\rm sgn} (V-u)= \theta (V-u)-\theta (u-V)$.
The relations (\ref{14}), (\ref{20}) permit to determine the shape of the
stationary probability density $\Phi (V)$. We find

\begin{equation}
\Phi (V) = \frac{(V-<V>)}{2I(V)}\left[ \fiV - \FiV \right] \label{23}
\end{equation}
\[+\frac{1}{4}\left[ \fiV + \FiV \right] \left\{ T_{1}+T_{2} 
			 +\phantom{\frac{1}{T}}\right. \]
\[+ \left. \int \frac{du}{I^{2}(u)}{\rm sgn}(V-u)\left[ \frac{k_{B}}{m}
(\phi_{T_{2}}^{m}(u)-\phi_{T_{1}}^{m}(u)) + <V> \left( \int_{-\infty}^{u}dv\Fiv
 + \int_{u}^{\infty}dv\fiv \right)\right] \right\} \]

In order to complete the calculation of $\Phi(V)$ we have still to evaluate
the average piston velocity $<V>$ of the piston which shows in equation
(\ref{23}). This can be achieved from the normalization condition

\begin{equation}
\int dV \Phi (V) = 1 \label{24}
\end{equation}

To perform the calculation we group together the terms multiplying $<V>$,
and rewrite
equation (\ref{23}) in the form

\begin{equation}
K(V)<V> = L(V) \label{25}
\end{equation}

It turns out that the function $K(V)$ can be written as a derivative
\begin{eqnarray}
K(V) & = & \frac{\partial}{\partial V}
        \left[ \frac{1}{I(V)}\left( \int_{-\infty}^{V}dv\Fiv
 + \int_{V}^{\infty}dv\fiv\right) \right. \label{26} \\
{} & + & \left. \frac{1}{2}I'(V)\int \frac{du}{I^{2}(u)}{\rm sgn}(V-u)
\left( \int_{-\infty}^{u}dv \Fiv  + \int_{u}^{\infty}\fiv \right)\right]
\nonumber
\end{eqnarray}
We then deduce from (\ref{26}) the relation

\begin{equation}
\int_{-\infty}^{+\infty} dV K(V) = \frac{1}{2}\left( \frac{1}{T_{2}}-
 \frac{1}{T_{1}} \right)\int_{-\infty}^{+\infty}\frac{du}{I^{2}(u)}
\left( \int_{-\infty}^{u}dv \Fiv  + \int_{u}^{\infty}dv\fiv \right) \label{27}
\end{equation}
which follows directly from the asymptotic formulae (\ref{16}).
The right hand side of equation (\ref{25}) equals

\begin{equation}
L(V)= 2\Phi (V) - \frac{1}{2}( T_{1}+T_{2})\left[ \fiV + \FiV \right]
- \frac{k_{B}}{m}\frac{\partial}{\partial V}A(V) \label{28}
\end{equation}

where
\[ A(V)= \frac{1}{I(V)}[\phi^{m}_{T_{2}}(V)-\phi^{m}_{T_{1}}(V)]\ +\]
\[+ \frac{1}{2}\left( \ilV\Fiv - \igV\fiv \right) \int \frac{dv}{I^{2}(v)}
{\rm sgn}(V-v)
\left[ \phi^{m}_{T_{2}}(v)-\phi^{m}_{T_{1}}(v) \right] \]
Taking again into account the asymptotics (\ref{16}) and the normalization
(\ref{24}) we get

\begin{equation}
\int_{-\infty}^{+\infty} dV L(V) = -\frac{(T_{1}-T_{2})}{2T_{1}T_{2}}
\left[ T_{1}-T_{2}+\frac{k_{B}}{m}\int_{-\infty}^{+\infty}
\frac{dv}{I^{2}(v)}( \phi^{m}_{T_{2}}(v)-\phi^{m}_{T_{1}}(v)) \right]
\label{29}
\end{equation}

Integrating equation (\ref{25}) over the whole velocity space we thus
arrive at the following formula for the stationary mean velocity of the
piston

\begin{eqnarray}
<V>\;\;& = & \;\; \left\{ T_{2}-T_{1}+\frac{k_{B}}{m}\int_{-\infty}^{+\infty}
\frac{dv}{I^{2}(v)}[ \phi^{m}_{T_{1}}(v)-\phi^{m}_{T_{2}}(v)) \right\}
 \label{30} \\
{} & \times & \left\{ \int_{-\infty}^{+\infty}\frac{du}{I^{2}(u)}
\left[ \int_{-\infty}^{u}dv \Fiv + \int_{u}^{\infty}dv\fiv \right]
\right\}^{-1} \nonumber
\end{eqnarray}

The formula (\ref{30}) is the main result of our analysis. Whereas the mean
velocity vanishes for $T_{1}=T_{2}$, the piston acquires
a macroscopic systematic motion when $T_{1} \neq T_{2}$, and this under the
equal
pressures condition (\ref{2}).

In order to further investigate this phenomenon let us analyze the case of
a small temperature difference. We thus put
\[ T_{1}=T,\;\; T_{2}=T+\delta T, \;\; {\rm with}\;\; 0<\frac{\delta
T}{T}\ll 1 \]
When $\delta T\to 0$, the denominator in equation (\ref{30}) takes the
asymptotic form

\begin{equation}
T\sqrt{\frac{m}{k_{B}T}}\int dw h(w) +{\rm O}(\delta T) \label{31}
\end{equation}
where

\begin{equation}
 h(w) = \left[ \int dw|u-w|\phi (w)\right]^{-2}  \label{32}
\end{equation}
and $\phi (w)$ is the dimensionless Maxwell distribution
\[ \phi (w)= \frac{1}{\sqrt{2\pi}} {\rm exp}(-w^{2}/2) \]
On the other hand the numerator equals

\begin{equation}
\delta T \left[ 1-\frac{1}{2}\int dw h(w)(w^{2}-1)\phi (w)\right]
+ {\rm O}((\delta T)^{2}) \label{33}
\end{equation}

So, when $\delta T \to 0$, the mean velocity of the piston is given by

\begin{equation}
<V> = \sqrt{\frac{k_{B}T}{m}}\left(\frac{\delta T}{T}\right)
\left[ 1-\frac{1}{2}\int dw h(w)(w^{2}-1)\phi (w)\right]\left[
\int dw h(w)\right]^{-1} + {\rm O}(\delta T) \label{34}
\end{equation}
Integration by parts yields the relation
\[ \int dw  h(w)(w^{2}-1)\phi (w) = \int dw\, w\phi (w)h'(w)\]
The inequality

\begin{equation}
wh'(w)=-2[h(w)]^{3/2}w\int dw' {\rm sign}(w-w')\phi (w')\leq 0 \label{35}
\end{equation}
permits to determine the sign of the mean velocity of the piston. We find

\begin{equation}
<V>\;\; > \;\; 0,\qquad {\rm if} \;\; \delta T > 0. \label{37}
\end{equation}

The inequality (\ref{37}) shows that in the stationary state the piston
performs an uniform motion in the direction of the higher temperature
region. In fact, we checked by numerical analysis that the inequality
(\ref{37}) remained true for any $T_{2}>T_{1}$.

\section{Concluding comments}

We have studied the effect of fluctuations on the motion of the piston
separating two semi-infinite regions filled with perfect gases at equal
pressures.
The explicit stationary solution of the Boltzmann equation (\ref{23}),
(\ref{30}) obtained for the special case where the mass  of
the piston equals that of a fluid particle reduces to the equilibrium Maxwell
distribution when the temperatures on both sides of the piston are equal.
The mean velocity then vanishes. In fact, it has been rigorously shown
that the piston (tagged particle) performs in this case a clasical
diffusive motion \cite{jepsen65}, \cite{lebowitz67}.

When the temperatures are different, the space asymmetry induces asymmetry
of fluctuations which turns
out to have a macroscopic effect. The piston attains a stationary state
with a non-zero average velocity oriented toward the higher temperature
region. In view of the absence of macroscopic forces we discover here
a genuine molecular motor. There is some analogy between our prediction and
the "stochastic ratchet" problem where one observes that fluctuations
conspire with the spatial anisotropy to induce a macroscopic directed
motion in the absence of macroscopic forces \cite{abad98}, \cite{kohler98}.

We expect the same qualitative behavior in the case of a massive piston
whose mass is much larger than that of a gas molecule. Indeed,
when $M\gg m$ the Boltzmann equation, which was the basis of our analysis,
gives an adequate description of the dynamics because the creation of
dynamical
correlations between the piston and the gas molecules via the recollision
mechanism becomes negligible. The persistence of the non-vanishing average
velocity  in the $M\gg m$ regime at equal pressures and different
temperatures will be the subject of the forthcoming
article\cite{gruber99}. One can thus expect that the stochastic motion of the piston
as a whole can lead to an effective energy transfer between the
separated fluids in the absence of pressure difference. Hence, although the
piston has no internal degrees of freedom calling it adiabatic becomes a
misuse of the term.

It remains to be seen what is the relevance of our analysis to the
original adiabatic piston problem where the fluids occupy finite volumes.
This is a delicate question indeed, because in one dimension even for a
very massive piston the collisions with the boundaries will inevitably
induce recollisions: the piston
will interact back with the particles it had already met before. So, the
Boltzmann description cannot be reliable in the long time limit. Indeed,
at the time scale characterized by the time
necessary to cover the total length of the cylinder $(L_{1}+L_{2})$ with
the thermal velocity of the fluids the effects of dynamical
correlations will start to play an important role. However, for
sufficiently large volumes, after the mechanical equilibrium has been
established, one can expect (on a larger time scale) the occurence of a
net displacement of the piston, reducing the volume of the the higher
temperature fluid, as it is the case for the conducting piston.
It seems likely that for finite volumes the Boltzmann description can work
much better in two and three dimensions because then the recollision effects
due to boundaries become weaker. Anyway, for finite volumes one will have
to consider the system of kinetic equations coupling the evolution of the
state of the piston to that of the fluids.

\end{document}